\documentclass[twocolumn,showpacs,preprintnumbers,amsmath,amssymb,superscriptaddress]{revtex4-1}

\usepackage{graphicx}
\usepackage{dcolumn}
\usepackage{bm}
\usepackage{color}
\usepackage{ulem}

\newcommand*{\SAGA}{%
Department of Physics, Saga University, Saga, 840-8502 Japan}
\newcommand*{\PUSAN}{%
Department of Physics, Pusan National University, Busan, 609-735 Republic of Korea}
\newcommand*{\DUBNA}{%
Laboratory of Nuclear Problems, Joint Institute for Nuclear Research, 
Dubna, Moscow Region, 141980 Russia}
\newcommand*{\TAIWAN}{%
Department of Physics, National Taiwan University, Taipei, Taiwan 10617 Republic of China}
\newcommand*{\SOKENDAI}{%
Department of Particle and Nuclear Research, 
The Graduate University for Advanced Science (SOKENDAI), Tsukuba, Ibaraki, 305-0801 Japan}
\newcommand*{\KEK}{%
Institute of Particle and Nuclear Studies, 
High Energy Accelerator Research Organization (KEK), Tsukuba, Ibaraki, 305-0801 Japan}
\newcommand*{\OSAKA}{%
Department of Physics, Osaka University, Toyonaka, Osaka, 560-0043 Japan }
\newcommand*{\YAMAGATA}{%
Department of Physics, Yamagata University, Yamagata, 990-8560 Japan}
\newcommand*{\CHICAGO}{%
Enrico Fermi Institute, University of Chicago, Chicago, Illinois 60637, USA }
\newcommand*{\NDA}{%
Department of Applied Physics, National Defense Academy, Yokosuka, Kanagawa, 239-8686 Japan}
\newcommand*{\RCNP}{%
Research Center of Nuclear Physics, Osaka University, Ibaraki, Osaka, 567-0047 Japan}
\newcommand*{\KYOTO}{%
Department of Physics, Kyoto University, Kyoto, 606-8502 Japan}
\newcommand*{\IHEP}{%
Institute for High Energy Physics, Protvino, Moscow region, 142281 Russia}
\newcommand*{\GOMEL}{%
Scarina Gomel' State University, Gomel', BY-246699, Belarus}
\newcommand*{\ARIZONA}{%
Department of Physics and Astronomy, Arizona State University, Tempe, Arizona, USA}
\newcommand*{\RIKEN}{%
RIKEN SPring-8 Center, Sayo, Hyogo, 679-5148 Japan}
\newcommand*{\NY}{%
University  of Rochester, Rochester, NY 14627}
\newcommand*{\CERN}{%
CERN, CH-1211 Gen\`{e}ve 23, Switzerland}
\begin{document}
\title{Study of the $K^0_L \to \pi^0\pi^0\nu\bar{\nu}$ decay}


\author{R.~Ogata}\affiliation{\SAGA}
\author{S.~Suzuki}\affiliation{\SAGA}
\author{J.~K.~Ahn}\affiliation{\PUSAN} 
\author{Y.~Akune}\affiliation{\SAGA} 
\author{V.~Baranov}\affiliation{\DUBNA}
\author{K.~F.~Chen}\affiliation{\TAIWAN}
\author{J.~Comfort}\affiliation{\ARIZONA} 
\author{M.~Doroshenko}\altaffiliation{Present address: \DUBNA}\affiliation{\SOKENDAI} 
\author{Y.~Fujioka}\affiliation{\SAGA} 
\author{Y.~B.~Hsiung}\affiliation{\TAIWAN}
\author{T.~Inagaki}\affiliation{\SOKENDAI}\affiliation{\KEK} 
\author{S.~Ishibashi}\affiliation{\SAGA}
\author{N.~Ishihara}\affiliation{\KEK}
\author{H.~Ishii}\affiliation{\OSAKA} 
\author{E.~Iwai}\affiliation{\OSAKA}
\author{T.~Iwata}\affiliation{\YAMAGATA} 
\author{I.~Kato}\affiliation{\YAMAGATA} 
\author{S.~Kobayashi}\affiliation{\SAGA}
\author{S.~Komatsu}\affiliation{\OSAKA}
\author{T.~K.~Komatsubara}\affiliation{\KEK} 
\author{A.~S.~Kurilin}\affiliation{\DUBNA} 
\author{E.~Kuzmin}\affiliation{\DUBNA}
\author{A.~Lednev}\affiliation{\IHEP}\affiliation{\CHICAGO} 
\author{H.~S.~Lee}\affiliation{\PUSAN} 
\author{S.~Y.~Lee}\affiliation{\PUSAN} 
\author{G.~Y.~Lim}\affiliation{\KEK}
\author{J.~Ma}\affiliation{\CHICAGO}
\author{T.~Matsumura}\affiliation{\NDA}
\author{A.~Moisseenko}\affiliation{\DUBNA}
\author{H.~Morii}\altaffiliation{Present address: \KEK}\affiliation{\KYOTO}
\author{T.~Morimoto}\affiliation{\KEK}
\author{Y.~Nakajima}\affiliation{\KYOTO}
\author{T.~Nakano}\affiliation{\RCNP} 
\author{H.~Nanjo}\affiliation{\KYOTO}
\author{N.~Nishi}\affiliation{\OSAKA}
\author{J.~Nix}\affiliation{\CHICAGO}
\author{T.~Nomura}\altaffiliation{Present address: \KEK}\affiliation{\KYOTO}
\author{M.~Nomachi}\affiliation{\OSAKA}

\author{H.~Okuno}\affiliation{\KEK}
\author{K.~Omata}\affiliation{\KEK}
\author{G.~N.~Perdue}\altaffiliation{Present address: \NY}\affiliation{\CHICAGO}
\author{S.~Perov}\affiliation{\DUBNA}
\author{S.~Podolsky}\altaffiliation{Present address: \GOMEL}\affiliation{\DUBNA}
\author{S.~Porokhovoy}\affiliation{\DUBNA}
\author{K.~Sakashita}\altaffiliation{Present address: \KEK}\affiliation{\OSAKA} 
\author{T.~Sasaki}\affiliation{\YAMAGATA} 
\author{N.~Sasao}\affiliation{\KYOTO}
\author{H.~Sato}\affiliation{\YAMAGATA}
\author{T.~Sato}\affiliation{\KEK}
\author{M.~Sekimoto}\affiliation{\KEK}
\author{T.~Shimogawa}\affiliation{\SAGA}
\author{T.~Shinkawa}\affiliation{\NDA}
\author{Y.~Stepanenko}\affiliation{\DUBNA}
\author{Y.~Sugaya}\affiliation{\OSAKA}
\author{A.~Sugiyama}\affiliation{\SAGA}
\author{T.~Sumida}\altaffiliation{Present address: \CERN}\affiliation{\KYOTO}
\author{Y.~Tajima}\affiliation{\YAMAGATA}
\author{S.~Takita}\affiliation{\YAMAGATA} 
\author{Z.~Tsamalaidze}\affiliation{\DUBNA}
\author{T.~Tsukamoto}\altaffiliation{Deceased.}\affiliation{\SAGA} 
\author{Y.~C.~Tung}\affiliation{\TAIWAN}
\author{Y.~W.~Wah}\affiliation{\CHICAGO}
\author{H.~Watanabe}\altaffiliation{Present address: \KEK}\affiliation{\CHICAGO}
\author{M.~L.~Wu}\affiliation{\TAIWAN}
\author{M.~Yamaga}\altaffiliation{Present address: \RIKEN}\affiliation{\KEK}
\author{T.~Yamanaka}\affiliation{\OSAKA}
\author{H.~Y.~Yoshida}\affiliation{\YAMAGATA}
\author{Y.~Yoshimura}\affiliation{\KEK}
\author{Y.~Zheng}\affiliation{\CHICAGO}

\collaboration{E391a Collaboration}\noaffiliation


\begin{abstract}
The rare decay $K^0_L \to \pi^0\pi^0\nu\bar{\nu}$ was studied 
with the E391a detector at the KEK 12-GeV proton synchrotron.  
Based on $9.4\times 10^9$ $K^0_L$ decays, an upper limit of 
$8.1\times 10^{-7}$ 
was obtained for the branching fraction 
at 
90$\%$ confidence level.  
We also set a limit on the 
$K^0_L \to \pi^0\pi^0X \; (X\to$ invisible particles) process; 
the limit on the branching fraction 
varied from 7.0$\times 10^{-7}$ to 4.0$\times 10^{-5}$ 
for the mass of $X$ ranging
from 50 MeV/$c^2$ to 200 MeV/$c^2$.  

\end{abstract}

\date{\today}

\maketitle


\begin{figure*} 
\includegraphics[height=4.8cm]{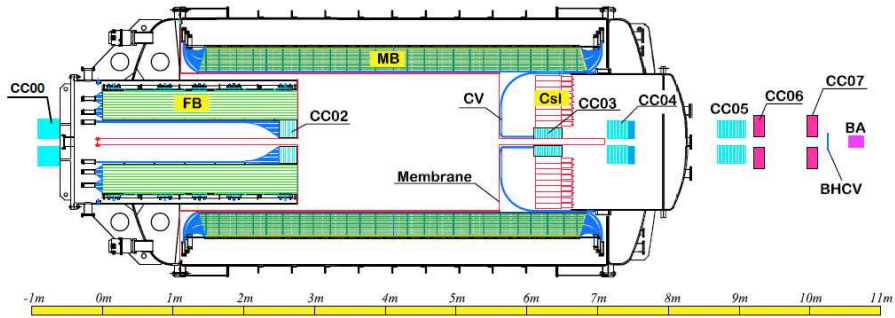}
   \caption{Schematic cross sectional view of the E391a detector.\
   The entrance of the detector is at ``0 m."} 
   \label{fig:detector}
\end{figure*}

\section{Introduction}
In the standard electroweak theory, flavor-changing neutral 
current (FCNC) processes are strongly suppressed
and can only occur via higher-order effects.  
Hence, these processes will be sensitive 
to unknown particles or interactions 
that contribute in higher-order loop diagrams.
Such processes are ideal places to look for 
signals of new physics beyond the Standard Model (SM).

In the SM, the FCNC $K^0_L \to \pi^0\pi^0\nu\bar{\nu}$ decay 
is predominantly a CP conserving process. 
Its branching fraction is sensitive to the real part of the 
$s\to d\nu\bar{\nu}$ transition amplitude, while the related 
decays $K^ 0_L\to\pi^0\nu\bar{\nu}$ and 
$K^+\to\pi^+\nu\bar{\nu}$ sense the imaginary part 
and absolute value, respectively. 
Like these decays, $K^0_L \to \pi^0\pi^0\nu\bar{\nu}$ is
theoretically clean and uncertainties in the hadronic matrix element 
can be removed by using the measured branching fraction of its relevant 
semileptonic decay $K^+\to\pi^0\pi^0e^+\nu$.

The SM predicts the branching fraction to be 
$(1.4\pm 0.4)\times 10^{-13}$ \cite{LV,ChingGilman}.
Although the prediction is solid in the SM, 
there is a possibility of enhancements from new physics 
contributions. 
In fact, phenomenological analyses give constraints 
on possible enhancements by up to an order of magnitude
within the allowed range of new physics parameters 
from known kaon decays, including the measured branching fraction
of $K^+\to\pi^+\nu\bar{\nu}$ \cite{ChingGilman}.

In addition, 
a new particle $X$ which decays into invisible particles, 
can also contribute to the three body decay $K^0_L \to \pi^0\pi^0X$. 
There is also a possibility of having new
pseudoscalar particles, as predicted by several 
supersymmetry models.  
Searches for the $X\to \gamma\gamma$ or $X\to \mu\bar{\mu}$ modes
and comparisons with the models have been made \cite{Xrefs}, and 
similar test should also be tried in the $\nu\bar{\nu}$ final state.

The E391a experiment at the KEK 12-GeV proton synchrotron made the first 
search for the $K^0_L \to \pi^0\pi^0\nu\bar{\nu}$ decay, 
based on the data sample taken in the first stage (Run-I).
An upper limit on the branching fraction of 4.7$\times 10^{-5}$ 
at the 90\% confidence level (C.L.) for 
$K^0_L\to\pi^0\pi^0\nu\bar{\nu}$ was obtained 
\cite{Nix}.  
Unfortunately the Run-I data were compromised by large neutron 
backgrounds from a vacuum membrane that hung in the beam. 
This article is based on the E391a data taken in the 
periods from February to April in 2005 (Run-II) and from November 
to December in 2005 (Run-III), after correcting the problem.  
These data had six times more $K^0_L$ 
decays and were analyzed with improved methods.   

\section{E391a detector}
The E391a detector was primarily designed to search for   
the $K^0_L \to \pi^0\nu\bar{\nu}$ decay \cite{e391paper,e391paper2}. 
It had a CsI calorimeter to detect two photons from a $\pi^0$ decay 
and hermetic veto counters to ensure that no other visible particles 
existed in the final state (Fig.\ \ref{fig:detector}). 
To avoid backgrounds from interactions of the beam particles with air, 
most of the detector components were placed in a vacuum vessel.

The CsI calorimeter consisted of 496 blocks of $7\times 7\times
30$ cm$^3$ undoped CsI crystal.  
A $12\times 12$ cm$^2$ beam hole 
was
located at the center of the 
calorimeter to allow the beam particles to pass through.  
The main barrel (MB) and front barrel (FB) counters were sandwiches 
of lead plates and plastic-scintillation counters with 13.5 $X_0$
and 17.5 $X_0$, respectively, and formed cylindrical walls surrounding 
the $K_L^0$ decay volume. 

Collar shaped counters (CC00, CC02$-$07) were placed around 
the beam line for vetoing photons along the beam axis. 
Charged particles that would hit the CsI calorimeter were identified 
and rejected by energy deposits in a charge veto (CV) 
scintillation-counter hodoscope, located 50 cm upstream; it covered 
the front face as well as the outer wall of the beam-hole area. 

Beam-hole charge veto (BHCV) and back-anti (BA) counters were 
located in the downstream region along the beam axis and outside 
of the vacuum vessel.  The BHCV consisted of eight 3-mm-thick
 plastic-scintillation counters, which were arranged to fully 
cover the downstream area of the beam hole.
The BA was made of six superlayers, each consisting of a 
lead/scintillator sandwich and quartz blocks for Run-II.  
For Run-III the BA had five superlayers, where 
the lead/scintillator sandwich part was replaced 
by the Lead Tungsten Oxide (PWO) crystals;  
this change was the only difference in the E391a detector system 
between the Run-II and Run-III periods.

Data acquisition was made with a hardware trigger that required 
two or more electromagnetic showers in the CsI calorimeter 
with total energy $>$ 60 MeV, and no obvious activity 
in the CV and other veto counters.
Loose on-line veto requirements 
were imposed, which ensured flexibility of setting 
energy thresholds consistently for the real data and 
the detector simulation during the off-line analysis.  
The $K^0_L\to \pi^0\pi^0 \nu \bar{\nu}$ decay should produce  
exactly four showers in the CsI calorimeter 
and nothing else in all the other counters. 
These hits should also satisfy an in-time requirement 
to suppress backgrounds.

\section{data reduction}
The experimental signature of $K^0_L \to \pi^0\pi^0\nu\bar{\nu}$ 
is four photons in the final states with a non-zero missing mass 
and a transverse momentum. 
Photons were registered in the CsI calorimeter as 
energy deposits in adjacent blocks (clusters). 
The most probable hit position of the photon was obtained from 
the shower energy distribution among the CsI blocks.  
Events with four photon clusters in the CsI were selected
for further off-line processing. 

Reconstruction of a pair of photons was made by assuming that 
they were from a $\pi^0$ that decayed on the beam axis ($z$-axis). 
Multiple pairings of four photons to reconstruct two $\pi^0$s 
were eliminated by taking the combination that gave the minimum 
$\chi^2$ value in the difference in $z$ vertices.
A cut on the minimum $\chi^2$ for the common $z$ vertex 
$V_z$ was effective for good $\pi^0$ identification. 
Also, requiring a large difference in the lowest two 
$\chi^2$ values was effective in reducing the combinatorial error.
We required that the common $\pi^0$-decay vertex point $V_z$
should be in the decay volume 300 cm $< V_z <$ 500 cm. 

The veto conditions on the MB, CV, and collar counters were 
carefully tuned. 
Among these counters, the MB played a major role in vetoing 
photons undetected by the CsI calorimeter.  
If too low an energy-threshold was imposed for all of the MB counters, 
back-splash from the CsI/CV surface and electromagnetic shower 
leakage from the outer part of the CsI calorimeter to the 
downstream area of the MB would cause acceptance loss.    
Thus, we applied a tighter threshold to the upstream region of the MB
to detect photons with a high efficiency, and 
a looser threshold to the downstream region of the MB to keep a large 
acceptance for the signal \cite{yuchen2}. 
Suppression of the fusion cluster (misidentification of two close-by 
clusters as one cluster) was made by using the neural
network (NN) technique trained by the stand-alone photons and 
the fusion clusters from the $K^0_L\to \pi^0\pi^0\pi^0$ Monte Carlo
(MC) simulation sample.
Further suppression was made on the shower shape of the cluster, 
and by examining the energy distribution and 
its RMS fraction \cite{yuchen2}.           

\begin{figure} [h]
  \includegraphics[height=8.5cm]{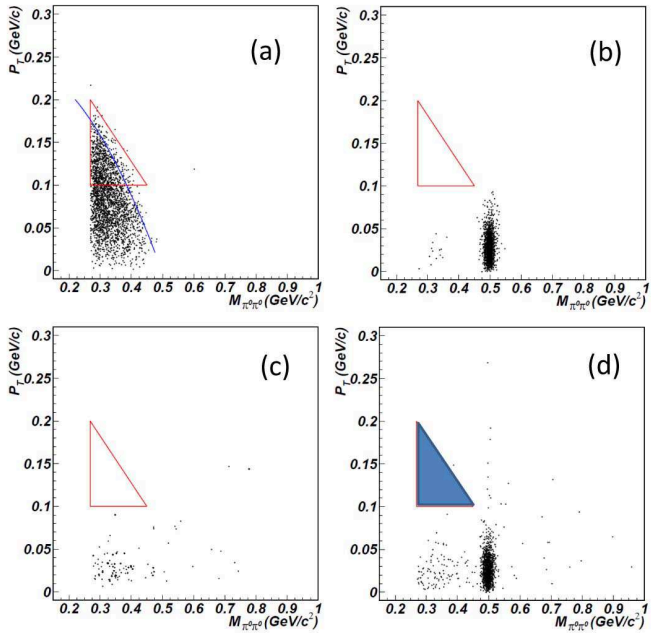}
   \caption{ $P_T$  vs. effective mass ($M_{\pi^0\pi^0}$) plots 
    of the $\pi^0\pi^0$ system from  Run-III MC simulations for 
    (a) $K^0_L\to\pi^0\pi^0\nu\bar{\nu}$, 
    (b) $K^0_L\to\pi^0\pi^0$, (c) $K^0_L\to \pi^0\pi^0\pi^0$, 
    and (d) Run-III real data. 
    The statistics of the MC events in (b) and (c) are normalized to 
    those of the real data. 
    The signal region is defined to be inside of the triangular area
    in the plots.  A curved 
    line in (a) indicates the kinematical limit for the 
    $K^0_L\to\pi^0\pi^0\nu\bar{\nu}$ decay.
   } 
  \label{fig:mcdata}
\end{figure}

\section{MC simulation}
A GEANT3-based \cite{geant3} MC simulation was performed 
to define the signal region for 
$K^0_L \to \pi^0\pi^0\nu\bar{\nu}$ and to estimate the acceptance. 
The $K^0_L \to \pi^0\pi^0\nu\bar{\nu}$ decay 
was generated with the V-A interaction, and the events were 
processed with the same reconstruction code as the real data.  
Among the incoming $K^0_L$ particles to the detector, 
2.66$\%$ decay within the fiducial decay volume.   
A total of $1\times 10^8$ incoming $K^0_L$ events were generated 
for both Run-II and Run-III.
The generated events were overlaid with accidental hits 
on the counters accumulated by special triggers 
during the data taking.
After applying the selection criteria described above, 
a two-dimensional distribution of the $\pi^0\pi^0$ effective mass 
$M_{\pi^0\pi^0}$ and the transverse momentum $P_T$ 
for the accepted events was examined.
The result 
of the simulation for $K_L^0\to\pi^0\pi^0\nu\bar{\nu}$
in the Run-III condition is shown in Fig.\ \ref{fig:mcdata}(a).

The signal region was defined as a right triangular area shown 
in Fig.\ \ref{fig:mcdata}. 
The vertical edge of the triangle is located at the lower bound of 
the $\pi^0\pi^0$ effective mass (0.268 GeV/$c^2$), and spanned the 
$P_T$ values between 0.1 to 0.2 GeV/$c$.
The horizontal 
edge defined the lower bound of $P_T$ (0.1 GeV/$c$) 
to avoid contamination from the $K^0_L \to \pi^0\pi^0\pi^0$ decay; 
it spanned the $M_{\pi^0\pi^0}$ values between 0.268 to 0.450 GeV/$c^2$.

The detector acceptance was calculated by taking the ratio of the number 
of events in the signal region to the number of $K^0_L$ 
particles that decayed in the fiducial vertex region.
A list of the reduction factors that contributed to the acceptance 
after the $K^0_L$s decayed in the fiducial volume are summarized in 
Table~\ref{tab:reduction}.
The acceptances obtained for Run-II and Run-III were consistent, and 
the result was $(3.02\pm0.11_{stat.})\times 10^{-4}$ for each run period.

\begin{table}
\caption{\label{tab:reduction}Reduction factors 
and resultant acceptances in each data reduction step (signal Monte Carlo).}
\begin{ruledtabular}
\begin{tabular} {lcl}
Cuts & Reduction   & Resultant \\
       & factor         & acceptance \\ 
\hline
$K_L^0$ decay in the fiducial volume &        &  1.00 \\
+ hardware trigger \& four-cluster- &  &  \\
~~~~~candidate in the CsI calorimeter         &  0.0499 & $4.99\times  10^{-2}$ \\
+ tighter CsI fiducial                 &  0.616 & $3.08\times  10^{-2}$ \\
+ rejection of fused cluster       & 0.361  & $1.11\times  10^{-2}$ \\
+ req. good cluster                  & 0.295  & $3.28\times  10^{-3}$ \\
+ $\chi^2$ cut for two $\pi^0$ vertex matching & 0.606  & $1.99\times  10^{-3}$ \\
+ other misc. requirements on CsI  & 0.829  & $1.65\times  10^{-3}$ \\
+ MB and upstream Vetoes       & 0.621  & $1.02\times  10^{-3}$ \\
+ downstream Vetoes              & 0.853   & $8.74\times  10^{-4}$ \\
+ req. signal box (after all the cuts) & 0.346  &$3.02\times  10^{-4}$ \\
\end{tabular}
\end{ruledtabular}
\end{table}

To study the dominant backgrounds, MC events of 
the $K^0_L \to \pi^0\pi^0$ decay with the statistics equivalent to 
11.2 and 11.9 times the Run-II and Run-III data, 
and the $K^0_L \to \pi^0\pi^0\pi^0$ decay equivalent to 0.80 and 2.52 
times the Run-II and Run-III data, respectively, 
were generated with the same detector conditions as the real data. 
Within these statistics, 
no MC events of these processes were found 
in the $K^0_L\to\pi^0 \pi^0 \nu\bar{\nu}$ signal region for both
the Run-II and Run-III conditions.

The MC results 
of the $K^0_L \to \pi^0\pi^0$ and $K^0_L \to \pi^0\pi^0\pi^0$ decays
for the Run-III condition, with the same statistics as the real data, 
are shown in Figs.\ \ref{fig:mcdata}(b) and (c), respectively, 
together with the Run-III real data (Fig.\ \ref{fig:mcdata}(d)).
The $K^0_L \to \pi^0\pi^0$ plot of $P_T$ vs.\ $M_{\pi^0\pi^0}$ 
showed clear clustering well removed from the signal region of 
$K^0_L \to \pi^0\pi^0\nu\bar{\nu}$ (Fig.\ \ref{fig:mcdata}(b)).
Events from the $K^0_L \to \pi^0\pi^0\pi^0$ decay were spread 
over the $P_T$ vs. $M_{\pi^0\pi^0}$ plot 
without having clear boundaries (Fig.\ \ref{fig:mcdata}(c)).
The data plot (Fig.\ \ref{fig:mcdata}(d)) outside of the 
masked signal region was characterized by the sum of 
Fig.\ \ref{fig:mcdata}(b) and (c),
except for a small number of events that extended to higher 
$P_T$ from the $K^0_L\to \pi^0\pi^0$ cluster region.  
These latter events are considered to be due to $K^0_S$ regeneration 
in the detector, and are not simulated in the MC.  They will 
be discussed in Sec.\ VI.
The MC study showed that most events stayed  
close to the signal region except that those for 
$M_{\pi^0\pi^0}\approx M_{K^0}$ were from 
$K^0_L \to \pi^0\pi^0$ and $K^0_L \to \pi^0\pi^0\pi^0$ decays. 
The main background source was $K^0_L \to \pi^0\pi^0\pi^0$.

\section{Sensitivity}
The number of $K_L^0$ particles that decayed in the fiducial 
vertex region was estimated by detecting $K_L^0 \to \pi^0\pi^0$ 
with the same $\pi^0$ selection criteria in the CsI calorimeter 
as for $K_L^0 \to \pi^0\pi^0\nu\bar{\nu}$.
Comparison with the MC simulation 
of $K_L^0\to \pi^0\pi^0$, with the 
same $\pi^0$ selection criteria as the real data, 
along with overlaid accidental events, 
gave the number of $K_L^0$ decays as 
$(5.48\pm0.09_{stat.}\pm0.31_{sys.})\times 10^9$ 
for Run-II and 
$(3.88\pm0.08_{stat.}\pm0.21_{sys.})\times 10^9$
for Run-III, respectively.

To estimate the systematic uncertainties, the fractional 
difference between the data and the MC simulation in each
selection criteria was examined, and the quadratic sum 
weighted by the effectiveness of each of the acceptance 
determinations \cite{e391paper2} was taken. 
The calculated values were 5.6\% and 5.5\% 
for Run-II and Run-III, respectively. 
The same values of the systematic uncertainty were 
adopted to the $K^0_L\to\pi^0\pi^0\nu\bar{\nu}$ decay. 

Based on the acceptance 
for the $K^0_L\to\pi^0\pi^0\nu\bar{\nu}$ decay
obtained from the MC simulation, 
and the number of $K^0_L$ decays in the data-taking runs, 
the single event sensitivity (S.E.S) was defined as
$$
S.E.S.=\frac{1}{Acceptance \times Number~of~K^0_L~decays}.
$$
The S.E.S. was derived as 
$(6.04\pm0.24_{stat.}\pm0.48_{sys.})\times 10^{-7}$
for Run-II and  
$(8.53\pm0.35_{stat.}\pm0.66_{sys.})\times 10^{-7}$ for Run-III.
The combined S.E.S. was 
$(3.54\pm0.10_{stat.}\pm0.28_{sys.})\times 10^{-7}$.
The sensitivity reported here is considerably improved from 
that of our previous analysis \cite{Nix}. 
In addition to the six-times larger statistics, the acceptance 
factor was improved for several selection criteria.  
Examples include the adoption of a looser cut for $\chi^2$ 
in the two-$\pi^0$ vertex matching, and optimizing the criteria  
used to select good photon clusters in the CsI calorimeter, 
both of which were possible in a 
cleaner environment with less neutron background
in Run-II and Run-III.

\begin{figure} 
  \includegraphics[height=5.4cm]{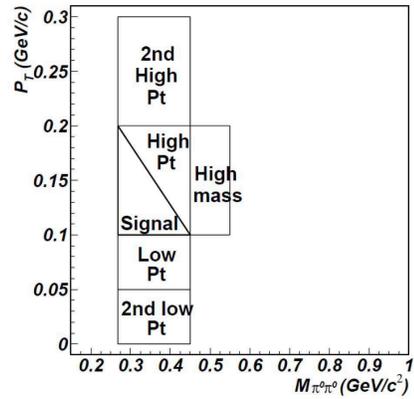}
   \caption{Kinematic boundaries of  $P_T$ vs. $M_{\pi^0\pi^0}$ for 
    the bifurcation analysis.} 
   \label{fig:bif-region}
\end{figure}

\begin{table}
\caption{\label{tab:estBGrun}Numbers of the events in the real data 
  (Run-II and Run-III combined) selected by the combinations 
  of two cut sets, A and B, at various regions defined 
  in Fig.\ \ref{fig:bif-region}.
  They were used for the estimations of $N_{bkg}$ in the bifurcation 
  method.  $N_{AB}$ is the observed number of events in those regions.
  }
\begin{ruledtabular}
\begin{tabular} {lrrrrrrr}
Region & $N_{A\bar{B}}$ & $N_{\bar{A}B}$ & $N_{\bar{A}\bar{B}}$ & 
Est. $N_{bkg}$ & $N_{AB}$ \\ 
\hline
Signal & 3 & 381 & 2508 & 0.46$\pm$0.26 & -- \\
\hline
Low $P_T$ & 136 & 6330 & 41418 & 20.8$\pm$1.8 & 21 & \\
2nd low $P_T$ & 1151 & 17525 & 105455 & 191.3$\pm$5.9 & 229 \\
High $P_T$ & 0 & 110 & 604 & 0.0 & 1 \\
2nd high $P_T$ & 1 & 5  & 41 & 0.12$\pm$0.13 & 0 \\
High Mass & 6 & 283  & 1086 & 1.56$\pm$0.65 & 12 \\

\end{tabular}
\end{ruledtabular}
\end{table}

\section{Background Estimation}
A ``blind analysis" technique was used to 
minimize the contribution from backgrounds without 
examining the candidate events in the signal region.  
Among the background sources, 
$K^0_L\to\pi^0\pi^0$ was fully reconstructed and rejected primarily 
by kinematic variables ($P_T$, $M_{\pi^0\pi^0}$) of the $\pi^0\pi^0$ 
system.  However, $K^0_L\to\pi^0\pi^0\pi^0$ would become a background
if two photons were missed due to finite photon-veto inefficiencies 
in the detector. 
The kinematic cuts could not be applicable to this dominant 
background source, and suppression depended strongly 
on the photon veto conditions.

To avoid the difficulty of having enough statistics 
in the MC simulation of the $K^0_L \to \pi^0\pi^0\pi^0$ 
background in the signal region, we adopted 
the ``bifurcation method" to estimate the number of background 
events in the signal region from the real data \cite{Nix,BNL}. 
We imposed several selection criteria (cuts) to identify 
the $K_L^0\to \pi^0\pi^0\nu\bar{\nu}$ signal from backgrounds,  
and the roles of these cuts were categorized in two groups, 
namely cut $A$ and cut $B$.
The cut $A$ consisted of ``veto" requirements to ensure
no particles other than from the CsI-calorimeter region, 
and the cut $B$ consisted of the 
cuts to select well-reconstructed $\pi^0$s from pairs of photon
clusters in the calorimeter.

If the two categories of cuts are uncorrelated, 
the number of events that passed both of the cuts $A$ and $B$ 
would be described as
$$N_{AB}\equiv N_0P(AB) \cong N_0P(A)P(B),$$ 
where $N_0$ is the number of events after the basic ``setup cuts"
are imposed prior to the cuts $A$ and $B$, and 
$P(A)$, $P(B)$ and $P(AB)$ are the probability of passing the 
conditions $A$, $B$, and $AB$, respectively.
The number of estimated background events $N_{bkg}$ is derived as 
\begin{eqnarray*}
N_{bkg}&=&N_0P(A)P(B) \\&=&
\frac{N_0P(A)P(\bar{B})N_0P(\bar{A})P(B)}{N_0P(\bar{A})P(\bar{B})} \\
&=&\frac{N_0P(A\bar{B})N_0P(\bar{A}B)}{N_0P(\bar{A}\bar{B})} 
=\frac{N_{A\bar{B}}N_{\bar{A}B}}{N_{\bar{A}\bar{B}}},
\end{eqnarray*}
where the symbols $\bar{A}$ and $\bar{B}$ are the inverse 
logic of $A$ and $B$, respectively.
Thus $N_{bkg}$ in the signal region 
after imposing all of the cuts $A$ and $B$ 
could be estimated from the number of events obtained by  
the combination of $A\bar{B}$, $\bar{A}B$, and $\bar{A}\bar{B}$.    

The kinematic region defined by $P_T$ and $M_{\pi^0\pi^0}$ of
the $\pi^0\pi^0$ system was further divided into six 
as shown in Fig.\ \ref{fig:bif-region}.
For each region, $N_{bkg}$ was estimated by the formula above.

\begin{figure}
  \includegraphics[height=4.2cm]{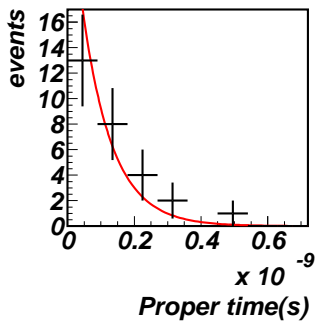}
   \caption{Proper-time distribution of the 
    $\pi^0\pi^0$ system for the events with $P_T\ge$ 0.1 GeV/$c$ 
    and $0.45\le M_{\pi^0\pi^0}\le$ 0.55 GeV/$c^2$ with a loose 
    requirement for $V_Z$ for the Run-II and Run-III, assuming 
    the production point was on the downstream edge of CC02.
    The solid line is an expectation from the 
    $K_S^0\to\pi^0\pi^0$ decay.}
   \label{fig:propertime}
\end{figure}

Table \ref{tab:estBGrun} summarizes the background estimates 
for the Run-II and Run-III data combined. 
Various kinematic regions were tabulated 
to cross-check the validity of this method.
The numbers of events of the real data $N_{AB}$
are consistent with the estimated values $N_{bkg}$, and also 
agree reasonably with the MC results of dominant backgrounds
(sum of $K^0_L\to\pi^0\pi^0$ and $\pi^0\pi^0\pi^0$ decays),
after imposing the cuts $A$ and $B$ (Table~\ref{tab:MCbifurcation})
in most of the regions around the signal region. 

The real data (12 events) in the ``High mass" region exceeded 
the estimated number of events ($1.56\pm 0.65$ events).
These events are seen in Fig.\ \ref{fig:mcdata}(d)
with $M_{\pi^0\pi^0}$ around $M_{K^0}$, extending to 
the higher $P_T$ region ($\ge$0.1 GeV/$c$). 
Fig.\ \ref{fig:propertime} shows the proper-time distribution
of the $\pi^0\pi^0$ system with $P_T \ge$ 0.1 GeV/$c$, 
$0.45 \le M_{\pi^0\pi^0} \le$ 0.55 GeV/$c^2$, and a looser 
$V_z$ requirement for the Run-II and Run-III combined, 
assuming that the production point was the downstream 
edge of the CC02 counter.   
The proper-time distribution (Fig.\ \ref{fig:propertime}) is 
consistent with that of expected from the $K^0_S\to\pi^0\pi^0$ 
decay, implying that the events 
are from regeneration of $K^0_S$ in the upstream materials
and subsequent $\pi^0\pi^0$ decay.  

Because the final state of the $K^0_S \to \pi^0\pi^0$ process 
was fully reconstructed, these events were insensitive to the 
photon-veto cuts (cuts $A$) and also to the CsI cluster shape cuts 
(cuts $B$). 
Hence the bifurcation method was not applicable for estimating 
the backgrounds from this process.
Nevertheless, these events might occur in the signal region 
by a mis-pairing of the photons from the $\pi^0$s.
Assuming that the mis-pairing rate for $K^0_S\to\pi^0\pi^0$
decays was equal to that for $K^0_L\to\pi^0\pi^0$ decays shown
in Fig.\ \ref{fig:mcdata}(b), the contribution of events 
in the ``High mass" region to the signal region was 
negligibly small ($<$ 0.03 events). 

The number of background events in the signal region
was estimated to be 0.46$\pm$0.26 for 
the Run-II and Run-III combined (Table~\ref{tab:estBGrun}).

\begin{table}
\caption{\label{tab:MCbifurcation}Comparison of MC events
of background processes
($K_L^0\to\pi^0\pi^0$ and $K_L^0\to\pi^0\pi^0\pi^0)$)
after imposing cuts $A$ and $B$ 
and real data in the kinematic regions given in
table~\ref{tab:estBGrun}.  
The MC results 
are normalized to the $K_L^0$ flux for the real data.
  }
\begin{ruledtabular}
\begin{tabular} {lrrrrrrr}
       & $K_L^0\to\pi^0\pi^0$ & $K_L^0\to\pi^0\pi^0\pi^0$ & Sum of & Real \\
Region & (MC)         & (MC)              & the two MC    & data \\ 
\hline
Signal        &  0.0 &   0.0 &   0.0 & --  \\
\hline
Low $P_T$     &  0.5 &  15.6 &  16.1 &  21 \\
2nd low $P_T$ & 11.9 & 203.1 & 215.0 & 229 \\
High $P_T$    &  0.1 &   0.0 &   0.1 &   1 \\
2nd high $P_T$ & 0.0 &   0.0 &   0.0 &   0 \\
High Mass     &  0.2 &   0.0 &   0.2 &  12 \\
\end{tabular}
\end{ruledtabular}
\end{table}

In the discussions above, we assumed that 
the cut sets $A$ and $B$ were uncorrelated.
A measure of the correlation was given  
by the parameter $\epsilon \equiv P(A|B)-P(A|\bar{B})$,
as the difference of the probability of satisfying the cut $A$
for the events that passed the cut $B$, and 
for the events that passed the cut $\bar{B}$.
Corrections to the $N_{bkg}$ due to the correlation are given by    
$$C_{\epsilon}=\epsilon\times N_{\bar{A}B}
\left( 1+\frac{N_{bkg}}{N_{\bar{A}B}}\right) $$
to first order of $\epsilon$ \cite{Nix,Nix2}.   
In the ``Low $P_T$" region, which was closest to the signal region,
$\epsilon$ was $(0.34\pm7.73)\times 10^{-4}$. 
Applying this value to the signal region, the correction to be made 
to the $N_{bkg}$ was $0.01\pm 0.29$ events.
The magnitude of the correction is small compared with 
the quoted error for the $N_{bkg}$, and this effect was not 
taken into account in deriving the final result.

\begin{figure}
  \includegraphics[height=5.7cm]{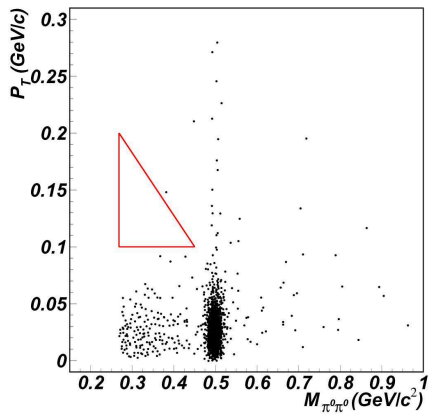}
   \caption{$P_T$ vs. $M_{\pi^0\pi^0}$ distribution for 
   Run-II and  Run-III combined. 
   } 
   \label{fig:openData}
\end{figure}

\section{Results}
After all of these studies, the candidate events 
in the signal region were examined.
Figure \ref{fig:openData} shows the combined  
results of the Run-II and Run-III; 
no events are observed in the signal region.
By using Poisson statistics
with zero background events in the signal region, 
we set an upper limit on the branching fraction of 
$K^0_L \to \pi^0\pi^0\nu\bar{\nu}$ to be 
8.1$\times 10^{-7}$ at the 90\% C.L.
The systematic uncertainty in the S.E.S.
is not taken into account in the limit.

From the same data sample, 
we also derived the upper limit on the
$K^0_L \to \pi^0\pi^0X$ ($X\to$ invisible particles) decay
assuming three-body phase space.
Because the acceptance depended on 
the mass of $X$, the upper limit 
was obtained as a function of the assumed mass of $X$ 
as summarized in Table \ref{tab:sensitivity}.

\begin{table}
\caption{\label{tab:sensitivity}Summary of the acceptance, S.E.S., and 
90$\%$ C.L. upper limit on the branching fraction
with several assumptions of the mass of X in the $K^0_L\to\pi^0\pi^0X$ 
($X\to$ invisible particles) decay. }
\begin{ruledtabular}
\begin{tabular} {lccc}
$M_{X}$       & accep-  &        & Br upper limit\\
(MeV/c$^2$)   & tance   & S.E.S. & with 90\% C.L.\\
\hline
50 & 3.52$\times 10^{-4}$ & 3.04$\times 10^{-7}$ & 7.0$\times 10^{-7}$ \\
75 & 3.50$\times 10^{-4}$ & 3.05$\times 10^{-7}$ & 7.0$\times 10^{-7}$ \\
100 & 3.45$\times 10^{-4}$  & 3.10$\times 10^{-7}$ & 7.1$\times 10^{-7}$ \\
125 & 2.91$\times 10^{-4}$  & 3.67$\times 10^{-7}$ & 8.5$\times 10^{-7}$ \\
150 & 2.07$\times 10^{-4}$  & 5.16$\times 10^{-7}$ & 1.2$\times 10^{-6}$ \\
175 & 9.18$\times 10^{-5}$  & 1.16$\times 10^{-6}$ & 2.7$\times 10^{-6}$ \\
200 & 6.21$\times 10^{-6}$  & 1.72$\times 10^{-5}$ & 4.0$\times 10^{-5}$ \\
\end{tabular}
\end{ruledtabular}
\end{table}

There are prospects of further improving the limit. 
A  new experiment E14 is now in preparation at J-PARC \cite{koto},
aiming at studying the $K^0_L\to\pi^0\nu\bar{\nu}$ decay
with three orders-of-magnitude higher sensitivity than E391a
by utilizing higher beam flux, longer beam time, and an 
upgraded detector with CsI calorimeter of higher efficiency and 
granularity. Further increases of the beam flux are planned in 
the upgrade.
In the case of $K^0_L\to \pi^0\pi^0\nu\bar{\nu}$, improving 
the photon cluster detection in the CsI calorimeter
is essential to improving the sensitivity. 
Hence we are hopeful to have significant progress on this  
process in the near future.     

\begin{acknowledgements}
We are grateful to the operating crew of the KEK 12-GeV proton 
synchrotron for the successful beam operation during the experiment.
This work has been partly supported by a Grant-in-Aid from the MEXT 
and JSPS in Japan, a grant from NSC and Ministry of Education in Taiwan, 
from the U.S. Department of Energy and from the Korea Research 
Foundation.  
\end{acknowledgements}

\bibliographystyle{physics}

\end{document}